# General individual attack on the ping–pong protocol with completely entangled pairs of qutrits

E. V. Vasiliu

Odesa national academy of telecommunication named after O.S. Popov,
Kovalska 1, 65029, Odesa, Ukraine

**Abstract.** *The general individual (non-coherent) attack on the ping–pong protocol with completely entangled pairs of three–dimensional quantum systems (qutrits) is analyzed. The expression for amount of the eavesdropper's information as functions from probability of attack detection is derived. It is shown, that the security of the ping–pong protocol with pairs of qutrits is higher the security of the protocol with pairs of qubits. It is also shown, that with the use by legitimate users in a control mode two mutually unbiased measuring bases the ping–pong protocol with pairs of qutrits, similar to the protocol with groups of qubits, possesses only asymptotic security and requires additional methods for its security amplification.*

## 1. INTRODUCTION

Now the quantum cryptography is one of the quickly developing applications of the quantum information theory and offers the new based on the quantum physics' laws approach to the solving of the important problem of telecommunication channels protection from eavesdropping by the not authorised persons. One of quantum cryptography directions is quantum secure direct communication protocols [1–8], in which a confidential message coded in quantum states is transmitted by a quantum communication channel without preliminary enciphering of the message.

One of the quantum secure direct communication protocol is the ping–pong protocol which can be realised with completely entangled states of pairs or groups of qubits. Some variants of the ping-pong protocol with qubits are developed, and also their security against various attacks is investigated [1, 2, 5–13]. Usage instead of qubits the quantum systems with high dimension will allow increasing information capacity of a source. So, the protocol with completely entangled three–dimensional quantum systems (qutrits) and quantum superdense coding for qutrits have the capacity $\log_2 9 = 3.17$ bits on a cycle instead of two bits on a cycle for the protocol with pairs of qubits. Notice, that from the technological point of view to operate with qutrits is more difficult so far than with qubits, however, a series of experiments on creation of entangled qutrit pairs is carried out by the present time [14, 15].

The quantum secure direct communication protocol with entangled pairs of qutrits, using the scheme of the ping–pong protocol, and also the transmition of qutrits by blocks was offered by Wang *et al.* [16]. However, such protocol with qutrits transmited by blocks requires great quantum memory from both users. At the same time the original scheme of the ping–pong protocol requires quantum memory for storage of one qubit (or qutrit) only from the receiver of the message during one cycle of the protocol. However, ping–pong protocol with groups of entangled qubits in Greenberger–Horne–Zeilinger (GHZ) states possesses only asymptotic security against general individual attack [1, 5, 6, 9, 10, 13]. For the protocol with qutrits such attack by this time is not analysed. In this paper the analysis of general individual attack against the ping–pong protocol with entangled pairs of qutrits is carried out; and also the comparison of security of this protocol with security of the protocols with qubits is fulfilled.

## 2. THE PING–PONG PROTOCOL WITH COMPLETELY ENTANGLED PAIRS OF QUTRITS

There are nine completely entangled orthonormalized states of qutrit pair (Bell states of qutrit pair) [16]:

$$|\Psi_{00}\rangle = (|00\rangle + |11\rangle + |22\rangle)/\sqrt{3}; \quad |\Psi_{10}\rangle = (|00\rangle + e^{2\pi i/3}|11\rangle + e^{4\pi i/3}|22\rangle)/\sqrt{3};$$
$$|\Psi_{20}\rangle = (|00\rangle + e^{4\pi i/3}|11\rangle + e^{2\pi i/3}|22\rangle)/\sqrt{3}; \quad |\Psi_{01}\rangle = (|01\rangle + |12\rangle + |20\rangle)/\sqrt{3};$$
$$|\Psi_{11}\rangle = (|01\rangle + e^{2\pi i/3}|12\rangle + e^{4\pi i/3}|20\rangle)/\sqrt{3}; \quad |\Psi_{21}\rangle = (|01\rangle + e^{4\pi i/3}|12\rangle + e^{2\pi i/3}|20\rangle)/\sqrt{3};$$
$$|\Psi_{02}\rangle = (|02\rangle + |10\rangle + |21\rangle)/\sqrt{3}; \quad |\Psi_{12}\rangle = (|02\rangle + e^{2\pi i/3}|10\rangle + e^{4\pi i/3}|21\rangle)/\sqrt{3};$$
$$|\Psi_{22}\rangle = (|02\rangle + e^{4\pi i/3}|10\rangle + e^{2\pi i/3}|21\rangle)/\sqrt{3}. \tag{1}$$

These states can be transformed one into another by the action of local unitary operations on one of qutrits from pair [16]. Thus, there is a possibility to realise quantum superdense coding for pair of qutrits, i.e. transmitting by a quantum communication channel only one qutrit from pair pass two classical trit of information.

The receiver (Bob) prepares one of two–qutrit states (1), let it be the state $|\Psi_{00}\rangle$, and then sends one of qutrits to the sender (Alice). Let Bob stores the first qutrit, the "home" qutrit and sends the second one, the "travel" qutrit.

Alice performs one of nine coding operations on the received qutrit, according to pair classical trits (ternary bigram) which she wishes to send. For example, the state $|\Psi_{00}\rangle$ corresponds to "00", $|\Psi_{10}\rangle$ corresponds to "10", $|\Psi_{20}\rangle$ corresponds to "20" etc. Alice and Bob agree about such conformity in advance. Then Alice sends the travel qutrit back to Bob, and he performs a Bell–basis measurement on both qutrits, representing a set from nine operators $\{|\Psi_{ij}\rangle\langle\Psi_{ij}|\}$, where $i, j = 0\ldots2$. Bob precisely defines a states created by Alice's coding operation, and, accordingly, ternary bigram which it has sent. The described sequence of operations is called as a message mode. Alice's coding operations, that transform state $|\Psi_{00}\rangle$ into states $|\Psi_{00}\rangle \ldots |\Psi_{22}\rangle$, are given by:

$$U_{00} = |0\rangle\langle 0| + |1\rangle\langle 1| + |2\rangle\langle 2|; \quad U_{10} = |0\rangle\langle 0| + e^{2\pi i/3}|1\rangle\langle 1| + e^{4\pi i/3}|2\rangle\langle 2|;$$
$$U_{20} = |0\rangle\langle 0| + e^{4\pi i/3}|1\rangle\langle 1| + e^{2\pi i/3}|2\rangle\langle 2|; \quad U_{01} = |1\rangle\langle 0| + |2\rangle\langle 1| + |0\rangle\langle 2|;$$
$$U_{11} = |1\rangle\langle 0| + e^{2\pi i/3}|2\rangle\langle 1| + e^{4\pi i/3}|0\rangle\langle 2|; \quad U_{21} = |1\rangle\langle 0| + e^{4\pi i/3}|2\rangle\langle 1| + e^{2\pi i/3}|0\rangle\langle 2|;$$
$$U_{02} = |2\rangle\langle 0| + |0\rangle\langle 1| + |1\rangle\langle 2|; \quad U_{12} = |2\rangle\langle 0| + e^{2\pi i/3}|0\rangle\langle 1| + e^{4\pi i/3}|1\rangle\langle 2|;$$
$$U_{22} = |2\rangle\langle 0| + e^{4\pi i/3}|0\rangle\langle 1| + e^{2\pi i/3}|1\rangle\langle 2|. \tag{2}$$

As the eavesdropper (Eve) has no access to home qutrit, stored in Bob quantum memory during one cycle of the protocol, she cannot gain any information having simply intercepted travel qutrit on the way Alice $\to$ Bob and having measured its state. The state of travel qutrit is completely mixed; its reduced density matrix is $\rho_{red} = \frac{1}{3}(|0\rangle\langle 0| + |1\rangle\langle 1| + |2\rangle\langle 2|)$. However, Eve has the possibility to perform an attack with the additional quantum systems (ancillas) entangled with travel qutrit on the way Bob $\to$ Alice (fig. 1). Then Eve performs measurement on the composed quantum system "travel qutrit – ancilla" on the way Alice $\to$ Bob (entanglement–and–measurement attack strategy). Therefore, except message mode in the ping–pong protocol it is also necessary control mode allowing to detect eavesdropper's operation.

Alice randomly switches to control mode with some probability $q$. In this mode Alice does not perform coding operations (2), but randomly chooses one of mutually unbiased (complementary) measuring basis, measures state of the travel qutrit in this basis and informs Bob via public classical channel a result of measurement and the chosen basis.

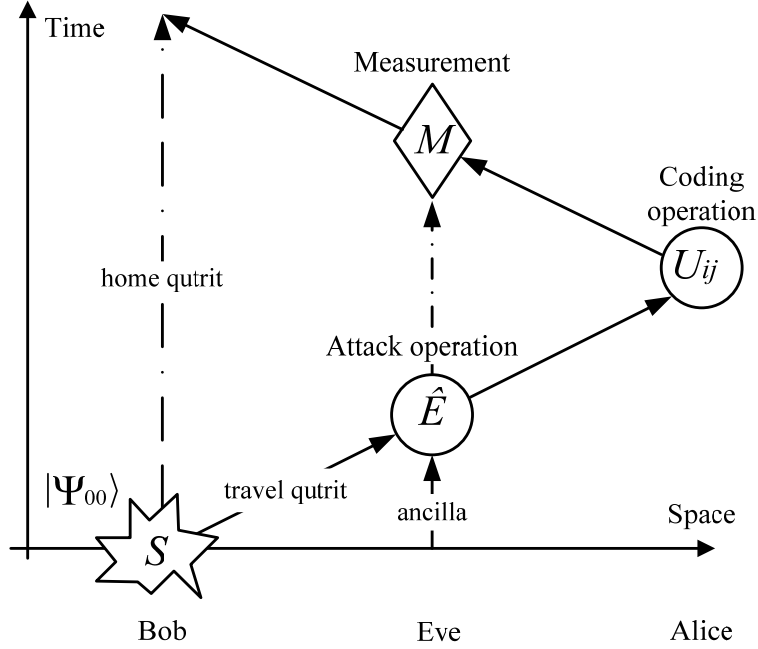

FIG. 1: The attack scheme on the ping–pong protocol with pairs of qutrits
($S$ is a source of entangled qutrit pairs)

There are four mutually unbiased bases for qutrits from which two are called $z$–basis and $x$–basis [16], and other two we will designate as $v$–basis and $t$–basis:

$$|z_0\rangle = |0\rangle, \qquad |z_1\rangle = |1\rangle, \qquad |z_2\rangle = |2\rangle; \qquad (3)$$

$$|x_0\rangle = (|0\rangle + |1\rangle + |2\rangle)/\sqrt{3},$$
$$|x_1\rangle = (|0\rangle + e^{2\pi i/3}|1\rangle + e^{-2\pi i/3}|2\rangle)/\sqrt{3},$$
$$|x_2\rangle = (|0\rangle + e^{-2\pi i/3}|1\rangle + e^{2\pi i/3}|2\rangle)/\sqrt{3}; \qquad (4)$$

$$|v_0\rangle = (e^{2\pi i/3}|0\rangle + |1\rangle + |2\rangle)/\sqrt{3},$$
$$|v_1\rangle = (|0\rangle + e^{2\pi i/3}|1\rangle + |2\rangle)/\sqrt{3},$$
$$|v_2\rangle = (|0\rangle + |1\rangle + e^{2\pi i/3}|2\rangle)/\sqrt{3}; \qquad (5)$$

$$|t_0\rangle = (e^{-2\pi i/3}|0\rangle + |1\rangle + |2\rangle)/\sqrt{3};$$
$$|t_1\rangle = (|0\rangle + e^{-2\pi i/3}|1\rangle + |2\rangle)/\sqrt{3};$$
$$|t_2\rangle = (|0\rangle + |1\rangle + e^{-2\pi i/3}|2\rangle)/\sqrt{3}. \qquad (6)$$

Measurement in any of these bases gives one of three possible results: "0", "1" or "2", each with probability equal to $1/3$. Having received from Alice the result of measurement and the chosen basis, Bob performs measurement of home qutrit state. Thus Bob should choose measuring basis according to the rules which follow from the form of $|\Psi_{00}\rangle$ in all four bases (3) – (6):

$$|\Psi_{00}\rangle = (|00\rangle + |11\rangle + |22\rangle)/\sqrt{3} = (|x_0 x_0\rangle + |x_1 x_2\rangle + |x_2 x_1\rangle)/\sqrt{3} =$$
$$= (|t_0 v_0\rangle + |t_1 v_1\rangle + |t_2 v_2\rangle)/\sqrt{3} = (|v_0 t_0\rangle + |v_1 t_1\rangle + |v_2 t_2\rangle)/\sqrt{3}. \qquad (7)$$

This expression is derived by direct calculation of projective operators' actions in bases (4) – (6) on state $|\Psi_{00}\rangle$, which has been written in computational basis (3), i.e. $|\Psi_{00}\rangle = (|00\rangle + |11\rangle + |22\rangle)/\sqrt{3}$.

From the formula (7) the rules follow for Bob measurement. So, if Alice has chosen *z*-basis and has drawn the result "0" Bob should choose also *z*-basis and its result with determinacy will be "0". Similarly, if Alice result at measurement in *z*–basis is "1" or "2" Bob also should draw "1" or "2" in this basis accordingly. If Alice has chosen *x*-basis, so Bob should choose this basis. At Alice results are "0", "1" or "2" (corresponding to the states $|x_0\rangle$, $|x_1\rangle$ and $|x_2\rangle$) Bob with determinacy will draw "0", "2" or "1" accordingly. If Alice will choose *v*-basis then according to the (7) Bob should choose *t*-basis, and on the contrary, and the results of Bob measurements depending on Alice's results also follow from the expression (7).

If Bob results differ from above, this means that the state $|\Psi_{00}\rangle$ is changed by the transmission of qutrit from Bob to Alice. It can be caused by two reasons: Eve's attack or noise in a quantum communication channel. In this paper I do not consider the implementation of the ping–pong protocol with entangled qutrits in the noise channel and I consider that legitimate users use the ideal quantum channel. In that case the discrepancy of Bob measurement result to the expected one means Eve's attack and legitimate users should interrupt a communication session immediately. However, for the ping–pong protocol with groups of entangled qubits Eve's attack does not lead to that Alice and Bob find out the change of the state prepared by Bob at once at the first control mode. In protocols with qubits Eve's attacking operation is found out for one round of the control mode only with some probability and legitimate users should perform a certain quantity of rounds to make probability of attack detection as much as close to unit [1, 5, 6, 9, 10]. For the ping–pong protocol with entangled pairs of qutrits the situation will be similar. A concrete number of control mode rounds necessary for attack detection with the probability set beforehand it is possible to find out having analysed the Eve's eavesdropping attack. Results of such an analysis are presented in the following section of the paper.

### 3. GENERAL INDIVIDUAL ATTACK USING QUANTUM ANCILLAS AGAINST THE PING–PONG PROTOCOL WITH ENTANGLED PAIRS OF QUTRITS

According to the scheme of the ping–pong protocol Alice informs Bob via the public channel on switching to control mode after reception of travel qutrit from Bob. Eve listening in this channel learns about switching to control mode after attacking operation $\hat{E}$, but before its final measurement (fig. 1). Hence, in this case Eve will not perform the measurement. Thus, legitimate users can detect only an attacking operation $\hat{E}$ on a line Bob $\rightarrow$ Alice.

Due to the Stinespring dilation theorem [17] Eve's attacking operation $\hat{E}$ can be realised by an unitary operator on a Hilbert space of ancillas $H_E$ which dimension satisfies the condition $\dim H_E \leq (\dim H_B)^2$, where $\dim H_B = 3$ is the dimension of travel qutrit Hilbert space. Thus, Eve can use for attack the ancillas consisting from one or two qutrits. Attack using two-qutrits ancillas is more general and accordingly stronger, therefore we will analyse this attack.

As the state of travel qutrit is completely mixed, so similarly to the ping–pong protocol with qubits [1, 5, 8] it is possible to consider, that Bob sends qutrit in one of the states $|0\rangle$, $|1\rangle$ or $|2\rangle$ with identical probability equal to $1/3$.

Thus, states of composite system "travel qutrit – Eve's ancilla" after an attack $\hat{E}$ can be written as

$$|\psi^{(0)}\rangle = \hat{E}|0,\varphi\rangle = \alpha_0|0,\varphi_{00}\rangle + \beta_0|1,\varphi_{01}\rangle + \gamma_0|2,\varphi_{02}\rangle;$$
$$|\psi^{(1)}\rangle = \hat{E}|1,\varphi\rangle = \alpha_1|0,\varphi_{10}\rangle + \beta_1|1,\varphi_{11}\rangle + \gamma_1|2,\varphi_{12}\rangle;$$
$$|\psi^{(2)}\rangle = \hat{E}|2,\varphi\rangle = \alpha_2|0,\varphi_{20}\rangle + \beta_2|1,\varphi_{21}\rangle + \gamma_2|2,\varphi_{22}\rangle, \qquad (8)$$

where $\{|\varphi_{ij}\rangle\}$ ($i, j = 0\ldots2$) is set of two-qutrit states of Eve's ancilla.

Matrix representation of Eve's attacking operation is

$$\hat{E} = \begin{bmatrix} \alpha_0 & \alpha_1 & \alpha_2 \\ \beta_0 & \beta_1 & \beta_2 \\ \gamma_0 & \gamma_1 & \gamma_2 \end{bmatrix} \qquad (9)$$

Since $\hat{E}$ has to be unitary the following relations between parameters of Eve's ancillas must be fulfilled:

$$\alpha_i^* \alpha_j + \beta_i^* \beta_j + \gamma_i^* \gamma_j = \delta_{ij}, \qquad (10)$$

where $\delta_{ij}$ is Kronecker delta, $i, j = 0 \ldots 2$.

Also for the reason that a state of travel qutrit is completely mixed the following relations must be fulfilled:

$$|\alpha_0|^2 = |\beta_1|^2 = |\gamma_2|^2; \quad |\alpha_1|^2 = |\beta_2|^2 = |\gamma_0|^2; \quad |\alpha_2|^2 = |\beta_0|^2 = |\gamma_1|^2. \qquad (11)$$

Let's consider at first the case where Bob "sends $|0\rangle$". In this case the state of system "travel qutrit – Eve's ancilla" after attack $\hat{E}$ becomes $|\psi^{(0)}\rangle$ (see (8)).

After the performance (by Alice) of coding operations $U_{00}$, $U_{10}$, $U_{20}$, $U_{01}$, ... (2) with frequencies $p_{00}$, $p_{10}$, $p_{20}$, $p_{01}$, ... respectively, the density operator of system "travel qutrit – Eve's ancilla" will look like:

$$\rho^{(0)} = \sum_{i,j=0}^{2} p_{ij} |\psi_{ij}^{(0)}\rangle\langle\psi_{ij}^{(0)}|, \qquad (12)$$

where

$$|\psi_{00}^{(0)}\rangle = U_{00}|\psi^{(0)}\rangle = \alpha_0|0,\varphi_{00}\rangle + \beta_0|1,\varphi_{01}\rangle + \gamma_0|2,\varphi_{02}\rangle,$$

$$|\psi_{10}^{(0)}\rangle = U_{10}|\psi^{(0)}\rangle = \alpha_0|0,\varphi_{00}\rangle + \beta_0 e^{2\pi i/3}|1,\varphi_{01}\rangle + \gamma_0 e^{4\pi i/3}|2,\varphi_{02}\rangle,$$

$$|\psi_{20}^{(0)}\rangle = U_{20}|\psi^{(0)}\rangle = \alpha_0|0,\varphi_{00}\rangle + \beta_0 e^{4\pi i/3}|1,\varphi_{01}\rangle + \gamma_0 e^{2\pi i/3}|2,\varphi_{02}\rangle,$$

$$|\psi_{01}^{(0)}\rangle = U_{20}|\psi^{(0)}\rangle = \alpha_0|1,\varphi_{00}\rangle + \beta_0|2,\varphi_{01}\rangle + \gamma_0|0,\varphi_{02}\rangle,$$

$$|\psi_{11}^{(0)}\rangle = U_{11}|\psi^{(0)}\rangle = \alpha_0|1,\varphi_{00}\rangle + \beta_0 e^{2\pi i/3}|2,\varphi_{01}\rangle + \gamma_0 e^{4\pi i/3}|0,\varphi_{02}\rangle,$$

$$|\psi_{21}^{(0)}\rangle = U_{21}|\psi^{(0)}\rangle = \alpha_0|1,\varphi_{00}\rangle + \beta_0 e^{4\pi i/3}|2,\varphi_{01}\rangle + \gamma_0 e^{2\pi i/3}|0,\varphi_{02}\rangle,$$

$$|\psi_{02}^{(0)}\rangle = U_{02}|\psi^{(0)}\rangle = \alpha_0|2,\varphi_{00}\rangle + \beta_0|0,\varphi_{01}\rangle + \gamma_0|1,\varphi_{02}\rangle,$$

$$|\psi_{12}^{(0)}\rangle = U_{12}|\psi^{(0)}\rangle = \alpha_0|2,\varphi_{00}\rangle + \beta_0 e^{2\pi i/3}|0,\varphi_{01}\rangle + \gamma_0 e^{4\pi i/3}|1,\varphi_{02}\rangle,$$

$$|\psi_{22}^{(0)}\rangle = U_{12}|\psi^{(0)}\rangle = \alpha_0|2,\varphi_{00}\rangle + \beta_0 e^{4\pi i/3}|0,\varphi_{01}\rangle + \gamma_0 e^{2\pi i/3}|1,\varphi_{02}\rangle. \qquad (13)$$

The maximal amount $I_0$ of classical information which is accessible to Eve after measurement on composite system "travel qutrit – ancilla" is defined by Holevo entropy [18]:

$$I_0 = S(\rho^{(0)}) - \sum_{i,j=0}^{2} p_{ij} S(\rho_{ij}^{(0)}) = S(\rho^{(0)}), \qquad (14)$$

where $\rho_{ij}^{(0)} = |\psi_{ij}^{(0)}\rangle\langle\psi_{ij}^{(0)}|$; $S$ is von Neumann entropy and all $S(\rho_{ij}^{(0)})$ are equal to zero, as the states (13) at the conditions (10) are pure. Thus,

$$I_0 = S(\rho^{(0)}) \equiv -Tr\{\rho^{(0)} \log_3 \rho^{(0)}\} = -\sum_i \lambda_i \log_3 \lambda_i \text{ (trit)}, \qquad (15)$$

where $\lambda_i$ are eigenvalues of the density operator $\rho^{(0)}$ (12).

The quantity of $I_0$ shows how much information Eve can gain after final measurement on composite system "travel qutrit – ancilla".

For finding of eigenvalues $\lambda_i$ density operator $\rho^{(0)}$ (12) has been written in a matrix kind in the following orthogonal basis:

$$\{|0,\varphi_{00}\rangle, |1,\varphi_{00}\rangle, |2,\varphi_{00}\rangle, |0,\varphi_{01}\rangle, |1,\varphi_{01}\rangle, |2,\varphi_{01}\rangle, |0,\varphi_{02}\rangle, |1,\varphi_{02}\rangle, |2,\varphi_{02}\rangle\}. \quad (16)$$

The deduced matrix has the size $9 \times 9$ and here is not shown so the resultant expression is cumbersome.

By means of Wolfram Mathematica 7 symbolic toolkit it has been found, that the equation on eigenvalues for this density matrix can be factorized to three cubic equations of a following kind:

$$\lambda^3 - (p_{00} + p_{10} + p_{20})\lambda^2 + 3(|\alpha_0|^2|\beta_0|^2 + |\alpha_0|^2|\gamma_0|^2 + |\beta_0|^2|\gamma_0|^2)(p_{00}p_{10} + p_{00}p_{20} + p_{10}p_{20})\lambda -$$
$$- 27|\alpha_0|^2|\beta_0|^2|\gamma_0|^2 p_{00}p_{10}p_{20} = 0;$$

$$\lambda^3 - (p_{01} + p_{11} + p_{21})\lambda^2 + 3(|\alpha_0|^2|\beta_0|^2 + |\alpha_0|^2|\gamma_0|^2 + |\beta_0|^2|\gamma_0|^2)(p_{01}p_{11} + p_{01}p_{21} + p_{11}p_{21})\lambda -$$
$$- 27|\alpha_0|^2|\beta_0|^2|\gamma_0|^2 p_{01}p_{11}p_{21} = 0;$$

$$\lambda^3 - (p_{02} + p_{12} + p_{22})\lambda^2 + 3(|\alpha_0|^2|\beta_0|^2 + |\alpha_0|^2|\gamma_0|^2 + |\beta_0|^2|\gamma_0|^2)(p_{02}p_{12} + p_{02}p_{22} + p_{12}p_{22})\lambda -$$
$$- 27|\alpha_0|^2|\beta_0|^2|\gamma_0|^2 p_{02}p_{12}p_{22} = 0. \quad (17)$$

Other cases in (8) are similarly considered, i.e. when Bob instead of $|0\rangle$ "sends $|1\rangle$ or $|2\rangle$". In these cases the eigenvalues of the density matrix $\rho^{(1)}$ and $\rho^{(2)}$ taking into account relations (11) are defined by the same equations (17).

As it follows from the first expression in (8), in the case when Bob "sends $|0\rangle$" and in control mode the measuring basis $z$ is used, a probability to detect an attack is:

$$d_z = |\beta_0|^2 + |\gamma_0|^2 = 1 - |\alpha_0|^2. \quad (18)$$

Similarly, if Bob "sends $|1\rangle$ or $|2\rangle$" then

$$d_z = |\alpha_1|^2 + |\gamma_1|^2 = 1 - |\beta_1|^2 = 1 - |\alpha_0|^2 \quad \text{and} \quad d_z = |\alpha_2|^2 + |\beta_2|^2 = 1 - |\gamma_2|^2 = 1 - |\alpha_0|^2 \quad (19)$$

accordingly, taking into account relations (11). Thus, the general expression for probability of attack detection with using $z$-basis in control mode is defined by (18).

Using the expression (18) from the equations (17) it is possible for a case of Eve's *symmetric* attack ($|\beta_0|^2 = |\gamma_0|^2 = d_z/2$) to exclude parameters $\alpha_0$, $\beta_0$ and $\gamma_0$ of ancillas having entered in the equations (17) probability of an attack detection $d_z$. It will finally allow to express Eve's amount of information $I_0$ (15) through $d_z$.

As at symmetric attack $|\alpha_0|^2|\beta_0|^2 = |\alpha_0|^2|\gamma_0|^2 = (1-d_z)\frac{d_z}{2}$, $|\beta_0|^2|\gamma_0|^2 = \frac{d_z^2}{4}$ and $|\alpha_0|^2|\beta_0|^2|\gamma_0|^2 = (1-d_z)\frac{d_z^2}{4}$, the equations (17) become:

$$\lambda^3 - (p_{00} + p_{10} + p_{20})\lambda^2 + 3\left(d_z - \frac{3}{4}d_z^2\right)(p_{00}p_{10} + p_{00}p_{20} + p_{10}p_{20})\lambda - \frac{27}{4}(d_z^2 - d_z^3)p_{00}p_{10}p_{20} = 0;$$

$$\lambda^3 - (p_{01} + p_{11} + p_{21})\lambda^2 + 3\left(d_z - \frac{3}{4}d_z^2\right)(p_{01}p_{11} + p_{01}p_{21} + p_{11}p_{21})\lambda - \frac{27}{4}(d_z^2 - d_z^3)p_{01}p_{11}p_{21} = 0;$$

$$\lambda^3 - (p_{02} + p_{12} + p_{22})\lambda^2 + 3\left(d_z - \frac{3}{4}d_z^2\right)(p_{02}p_{12} + p_{02}p_{22} + p_{12}p_{22})\lambda - \frac{27}{4}(d_z^2 - d_z^3)p_{02}p_{12}p_{22} = 0. \quad (20)$$

On the fig. 2 dependences of $I_0$ from $d_z$ are shown at Eve's symmetric attack and various values of frequencies $p_{00} \ldots p_{22}$ of Alice's coding operations (tab. 1). For acquisition of these dependences the equations (20) were solved numerically at some values of $p_{00} \ldots p_{22}$ and the obtained nine values $\lambda_1 \ldots \lambda_9$ were substituted in the equation (15).

Tab. 1. Frequencies $p_{00} \ldots p_{22}$ of ternary bigram and corresponding source entropy $H = -\sum_{i,j=0}^{2} p_{ij} \log_3 p_{ij}$ (trit/bigramm)

| Number of curve on fig. 2 | $p_{00}$ | $p_{10}$ | $p_{20}$ | $p_{01}$ | $p_{11}$ | $p_{21}$ | $p_{02}$ | $p_{12}$ | $p_{22}$ | $H$ |
|---|---|---|---|---|---|---|---|---|---|---|
| 1 | 1/9 | 1/9 | 1/9 | 1/9 | 1/9 | 1/9 | 1/9 | 1/9 | 1/9 | 2.000 |
| 2 | 1/6 | 1/9 | 1/18 | 1/6 | 1/18 | 1/9 | 1/18 | 1/9 | 1/6 | 1.921 |
| 3 | 2/9 | 0 | 2/9 | 0 | 2/9 | 0 | 2/9 | 0 | 1/9 | 1.439 |
| 4 | 0.4 | 0.1 | 0 | 0 | 0.4 | 0.1 | 0 | 0 | 0 | 1.086 |
| 5 | 2/3 | 0 | 0 | 0 | 1/3 | 0 | 0 | 0 | 0 | 0.579 |

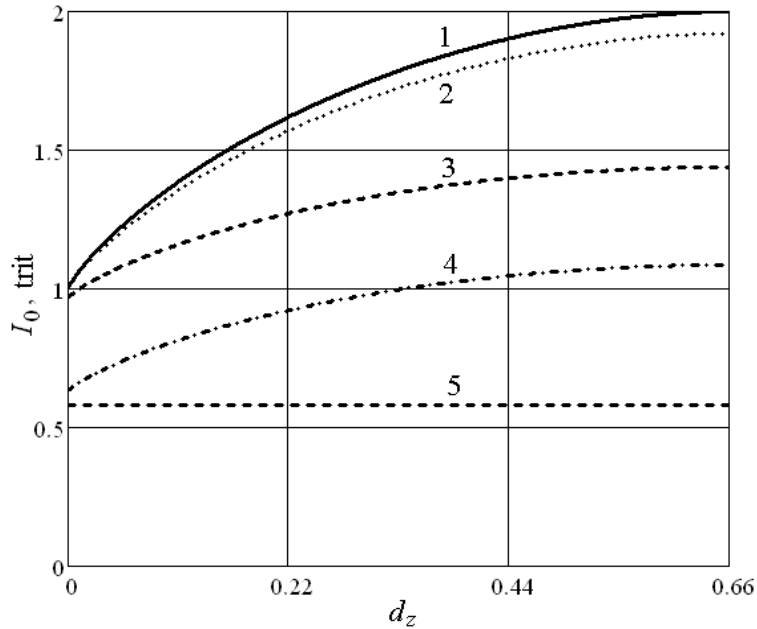

FIG. 2: Dependences of Eve's amount of information $I_0$ from probability $d_z$ of attack detection at symmetric attack

Apparently from the fig. 2, for the majority of sets $p_{00} \ldots p_{22}$ amount of Eve's information $I_0$ monotonously grows with increase of attack detection probability $d_z$ and reaches a maximum at $d_z = 2/3$ (for some special sets of bigram frequencies the $I_0$ does not depend on $d_z$ at all and is a constant). It is possible to consider the value $d_z = 2/3$ as a maximum so at $d_z > 2/3$ amount of Eve's information starts to decrease (on fig. 2 it is not shown). Accordingly, Eve will not choose parameters of the ancillas on which the $d_z$ depends so that $d_z$ would exceed $2/3$, since for her is not meaningful to increase probability of attack detection at reduction of information accessible to her. Also it is visible from the fig. 2 that a maximum of Eve's amount of information corresponding to $d_z = 2/3$ is equal to entropy of a message source at any sets of frequencies $p_{00} \ldots p_{22}$. It means that at $d_z = 2/3$ and only at such value of $d_z$ Eve have the full information. Also the fact that $I_0$ is equal to $H$ at $d_z = 2/3$ testifies about correct asymptotics of formulas (20).

It is visible from the fig. 2 also that at $d_z = 0$ Eve's amount of information does not equal to zero, however it is below its maximum value at $d_z = 2/3$. Thus, for the ping–pong protocol with entangled pairs of qutrits there is an "invisible" mode at which Eve obtaines the partial information, but its operations cannot be detected by legitimate users when they use only one measuring basis in control mode. Notice that the similar situation takes place and for the ping–pong protocol with entangled groups of qubits [5, 9, 10]. Therefore it is necessary to consider probabilities of attack detection at use by legitimate users in control mode other bases (4) – (6), and also dependences between these probabilities.

## 4. PROBABILITIES OF ATTACK DETECTION AT USE BY LEGITIMATE USERS $X$ – $V$ – AND $T$–BASES

Let's discuss Eve's attack considering that owing to full mixing of a travel qutrit state this qutrit is now in one of the states $|x_0\rangle$, $|x_1\rangle$ or $|x_2\rangle$ (4). Then formulas (8) are replaced with the following:

$$|\psi_x^{(0)}\rangle = \hat{E}|x_0,\varphi\rangle = a_0|x_0,\varphi_{00}\rangle + b_0|x_1,\varphi_{01}\rangle + c_0|x_2,\varphi_{02}\rangle;$$
$$|\psi_x^{(1)}\rangle = \hat{E}|x_1,\varphi\rangle = a_1|x_0,\varphi_{10}\rangle + b_1|x_1,\varphi_{11}\rangle + c_1|x_2,\varphi_{12}\rangle;$$
$$|\psi_x^{(2)}\rangle = \hat{E}|x_2,\varphi\rangle = a_2|x_0,\varphi_{20}\rangle + b_2|x_1,\varphi_{21}\rangle + c_2|x_2,\varphi_{22}\rangle. \quad (21)$$

Further, all formulas (9) – (17) remain fair at replacement $\alpha_0 \to a_0$, $\beta_0 \to b_0$, $\gamma_0 \to c_0$, $\alpha_1 \to a_1$, $\beta_1 \to b_1$ etc. Thus, expression (18) is replaced with expression

$$d_x = |b_0|^2 + |c_0|^2 = 1 - |a_0|^2. \quad (22)$$

Using (8) and (21) it is possible to derive the following expressions connecting parameters $\alpha_0$, $\beta_0$ and $\gamma_0$ with parameters $a_0$, $b_0$ and $c_0$:

$$|\alpha_0|^2 = \frac{1}{3}|a_0 + b_0 + c_0|^2, \quad |\beta_0|^2 = \frac{1}{3}|a_0 + e^{2\pi i/3}b_0 + e^{-2\pi i/3}c_0|^2, \quad |\gamma_0|^2 = \frac{1}{3}|a_0 + e^{-2\pi i/3}b_0 + e^{2\pi i/3}c_0|^2. \quad (23)$$

In the tab. 2 some calculated sets of parameters $a_0$, $b_0$ and $c_0$ satisfying to the relations similar to (10) and (11), and also corresponded to this parameters sets values of $d_x$ and $d_z$ are presented. Values of $d_z$ are obtained with the use of (18) and the first formula in (23).

Tab. 2. Parameters of attacking operation and corresponding to them values of $d_x$ and $d_z$

| $a_0$ | $b_0$ | $c_0$ | $d_x$ | $d_z$ |
|---|---|---|---|---|
| Nonsymmetric attack: $|b_0|^2 \neq |c_0|^2$ | | | | |
| –0.910684 | 0.244017 | –0.333333 | 0.170655 | 0.666667 |
| –0.807162 | 0.309719 | –0.502558 | 0.348490 | 0.666667 |
| –0.709081 | 0.331451 | –0.622370 | 0.497204 | 0.666667 |
| –0.666667 | 0.333333 | –0.666667 | 0.555556 | 0.666667 |
| –0.577406 | 0.325969 | –0.748563 | 0.666603 | 0.666667 |
| 0.530210 – 0.8i | 0.169304 – 0.1i | 0.014630 + 0.2i | 0.078878 | 0.666667 |
| –0.909127 + 0.1i | –0.133042 – 0.2i | 0.125653 – 0.3i | 0.163489 | 0.666667 |
| 0.204236 + 0.83i | 0.026660 – 0.3i | 0.136663 + 0.4i | 0.269388 | 0.666667 |
| 0.737034 + 0.3i | –0.031581 – 0.5i | 0.160573 – 0.3i | 0.366781 | 0.666667 |
| 0.674712 + 0.3i | 0.525520 + 0.2i | –0.220436 – 0.3i | 0.454764 | 0.666667 |
| –0.531662 + 0.3i | 0.463325 + 0.2i | –0.531662 + 0.3i | 0.627335 | 0.666667 |
| –0.497557+0.293i | 0.459068 + 0.2i | –0.570890 + 0.3i | 0.666667 | 0.666667 |

| Symmetric attack: $|b_0|^2 = |c_0|^2$ | | | | |
|---|---|---|---|---|
| –0.953939 + 0.1i | –0.2i | –0.2i | 0.08 | 0.666667 |
| 0.305505 + 0.8i | 0.305505 – 0.2i | 0.305505 – 0.2i | 0.266667 | 0.666667 |
| 0.027387 + 0.7i | 0.463276 – 0.2i | 0.463276 – 0.2i | 0.50925 | 0.666667 |
| 0.577350 | –0.288675 + 0.5i | –0.288675 + 0.5i | 0.666667 | 0.666667 |
| 0.577350i | 0.5 – 0.288675i | 0.5 – 0.288675i | 0.666667 | 0.666667 |
| 0.577350 | 0.288675 + 0.5i | 0.288675 + 0.5i | 0.666667 | 0.222222 |

The unitarity of Eve's attacking operation leads to the important dependence between $d_z$ and $d_x$, namely: without dependence from value of one of these quantities the second is always equal to the maximum value (see tab. 2). Thus, at use of two measuring bases in control mode an "invisible" mode does not exist any more and the ping–pong protocol with Bell pairs of entangled qutrits possesses asymptotic security like protocols with groups of entangled qubits [1, 5, 9, 10].

In a similar way the attack can be analysed at use by legitimate users of $v$–basis (5) and $t$–basis (6). The calculations show that the above–stated rules for $d_z$ and $d_x$ is fair for any of pairs $d_z - d_x$, $d_z - d_v$, $d_z - d_t$, $d_x - d_v$, $d_x - d_t$, $d_v - d_t$: without dependence from value of one of these quantities the second is always equal to the maximum value. From here follows that for eavesdropping detection it is enough for legitimate users to choose any two mutually unbiased bases from four possible.

At use in control mode two measuring bases (for example, $z$ and $x$) probability to detect Eve's attacking operation is

$$d = q_z d_z + q_x d_x, \qquad (24)$$

where $q_z$ and $q_x$ are probabilities of use by Alice and Bob $z$– and $x$–bases accordingly ($q_z + q_x = 1$). The least values of $d_z$ and $d_x$ are equal to zero but when one of these quaitities is equal to zero, another is equal to maximum value $2/3$. As legitimate users do not foreknow what attack strategy will be chosen by Eve, i.e. in what of bases she will aspire to create smaller value of detection probability, values $q_z$ and $q_x$ will be reasonable for choosing equal each other, i.e. $q_z = q_x = 1/2$. Then the least value of $d$ will turn out, when or $d_z = 0$ and $d_x = 2/3$, or on the contrary. According to (24) under such circumstances $d = (1/2) \cdot (2/3) = 1/3$, i.e. the minimum value of attack detection probability at use in control mode two measuring bases equals to $1/3$. Notice, that at such strategy Eve will obtain only partial information about the transmited string of trits. If Eve wants to obtain full information it should choose parameters of ancillas so that $d_z = d_x = 2/3$ and thus, according to (24), $d = 2/3$.

## 5. COMPARISON OF SECURITY LEVEL OF PROTOCOLS WITH QUTRITS AND WITH QUBITS

Let's now compare dependences of Eve's amount of information on probability of attack detection for protocol with Bell pairs of qutrits and protocols with groups of qubits. On fig. 3 dependences $I_0$ from $d_z$ for protocol with pairs of qutrits are shown at $p_{00} = \ldots = p_{22} = 1/9$, and also for protocol with pairs of qubits and quantum superdense coding [5], protocol with GHZ–triplets [9] and protocol with GHZ–quadruples of qubits [19] at identical values of frequencies of Alice's coding operations. It is visible, that the curve $I_0(d_z)$ for the protocol with qutrits lays close to corresponding curve for the protocol with GHZ–triplets of qubits. Also information capacities of these two variants of the ping–pong protocol are also close: 3.17 bits on a cycle for protocol with pairs of qutrits and 3 bits on a cycle for protocol with GHZ–triplets of qubits.

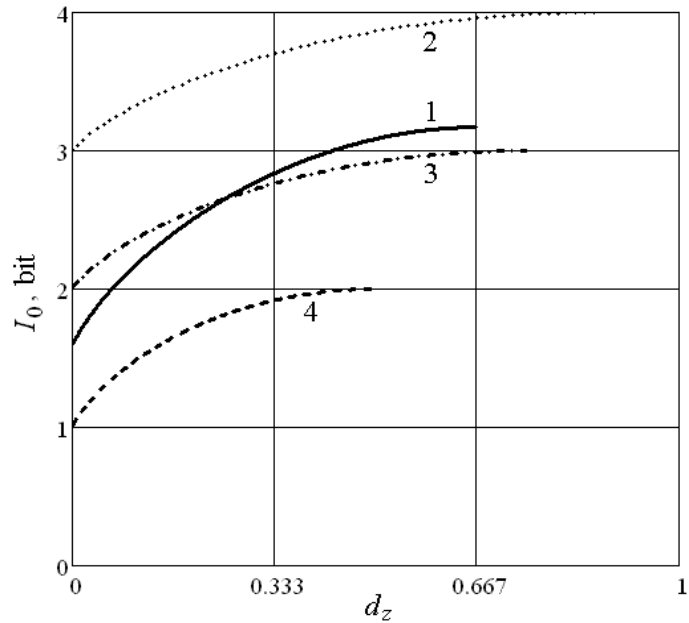

FIG. 3: Dependences of Eve's amount of information on probability of attack detection.
The protocol: with Bell pairs of qutrits (1); with GHZ–quadruples of qubits (2);
with GHZ–triplets of qubits (3); with Bell pairs of qubits (4).

In the tab. 3 the minimum and maximum values of attack's detection probability for these four variants of the ping–pong protocol are shown at use in control mode two measuring bases with the same probabilities $q_z = q_x = 1/2$. Apparently from the tab. 3, values $d_{min}$ and $d_{max}$ for protocol with Bell pairs of qutrits are also closest to corresponding values for protocol with GHZ–triplets of qubits.

Tab. 3. Minimum $d_{min}$ and maximum $d_{max}$ values of attack's detection probability at use in control mode two measuring bases with equal probabilities

| Protocol | $d_{min}$ | $d_{max}$ |
|---|---|---|
| with Bell pairs of qutrits | 1/3 | 2/3 |
| with Bell pairs of qubits | 1/4 | 1/2 |
| with GHZ–triplets of qubits | 3/8 | 3/4 |
| with GHZ–quadruples of qubits | 7/16 | 7/8 |

### 6. CONCLUSIONS

In this paper a general individual attack on the ping–pong protocol with Bell pairs of three–dimensional quantum systems is analysed. The density matrix of composite quantum system "travel qutrit – eavesdropper's ancilla" is obtained and by calculating of density matrix eigenvalues the expression is obtained for amount of eavesdropper's information as functions of attack's detection probability.

It is shown, that at use for protocol implementation Bell pairs of qutrits instead of Bell pairs of qubits not only information capacity increases, but also security level of protocol to attack, as the maximum probability of eavesdropping detection (at the one-time run of control mode) for protocol with qutrits is equal to $2/3$, and for protocol with qubits is equal to $1/2$. Security of the protocol with pairs of qutrits appears approximately same, as the protocol with GHZ–triplets of qubits. Also for protocol with pairs of qutrits, similar to the protocol with groups of qubits, there is an "invisible" mode if legitimate users use in control mode only one measuring basis. Use of the second measuring basis eliminates the possibility of undetectable attacks and, hence, is necessary.

At use in control mode two measuring bases the ping–pong protocol with Bell pairs of qutrits, similar to the protocol with groups of qubits, possesses only asymptotic security as for detection of eavesdropping with the probability as much as close to unit, it is necessary for legitimate users to perform a certain quantity of control mode rounds. Thus, as the control mode is necessary for alternating with the message mode (differently the eavesdropper will not make attacking operations at all as he will know that the message is not transmitting) some amount of information will leak to the eavesdropper. An estimation of this amount depending on parameters of protocol and eavesdropping strategy, and also necessary arrangements on security amplification of the ping–pong protocol with qutrits will be a subject of the following paper.